\documentclass[twocolumn]{webofc}

\usepackage[varg]{txfonts}   
\usepackage{hyperref}
\usepackage{url}
\usepackage[rightcaption]{sidecap}

\hypersetup{colorlinks=true,citecolor=blue,urlcolor=blue,linkcolor=blue}

\begin{document}
\title{Photoelastic Grain Solver v2.0: An updated tool for analysis of force measurements in granular materials}
%
%

\author{
        \firstname{Carmen L.} \lastname{Lee}\inst{1} \and
        \firstname{Lori} \lastname{McCabe}\inst{2} \and
        \firstname{Ben} \lastname{McMillan}\inst{3} \and
        \firstname{Abrar} \lastname{Naseer}\inst{4} \and
        \firstname{Dong} \lastname{Xie}\inst{5} \and
        \firstname{Ted} \lastname{Brzinski}\inst{6}\fnsep\thanks{\email{tbrzinski@haverford.edu}} \and
        \firstname{Karen E.} \lastname{Daniels}\inst{1} \and
        \firstname{Tejas} \lastname{Murthy}\inst{4} \and
        \firstname{Kerstin} \lastname{Nordstrom}\inst{2}
        }

\institute{Department of Physics, North Carolina State University, NC, USA 
\and
           Mount Holyoke College, MA, USA
\and
           University of Cambridge, UK
\and
           Indian Institute of Science
\and
        University of Edinburgh, UK
\and
           Haverford College, PA, USA
          }

\abstract{Photoelastic force imaging is an
experimental technique whereby a birefringent granular material is imaged with a polariscope to characterize the internal stress state of a granular material. Photoelasticimetry is the only proven experimental technique that allows researchers to measure the shear and normal forces at every particle contact in a granular packing. In 2017, Kollmer et al. [Rev. Sci. Instrum. 88, 051808 (2017)] developed an open-source software to perform this analysis. Here, we present the next substantial update to this software package. The new version improves resolution and efficiency and substantially changes the software architecture. The structural changes better facilitate add-ons, modules, and future improvements to the performance, accessibility, and versatility of the tool. Besides updates to the core software, we introduce new infrastructure to support the ongoing development of software, documentation, and training materials. The full development team, software, and supporting resources are available at \url{https://github.com/photoelasticity}. 
}

\maketitle

\section{Introduction \label{sec:intro}}

Our current best models for understanding the mechanics of granular materials
relate the bulk properties of the system to the internal state of the material, as encoded by the structures and interactions at all scales down to the particle scale.
The internal state of a granular material is challenging to characterize in experiments because the materials, in general, are optically opaque, and few techniques exist to assess the forces between particles, generally dominated by contact forces.
Researchers have been able to capture the internal structure of granular packings using techniques including index-matched fluorescence tomography~\cite{dijksman_refractive_2017} and x-ray micro computed-tomography ($\mu$-CT)~\cite{hall_discrete_2010}. While interparticle force measurements via $\mu$-CT are increasingly possible \cite{hurley_quantifying_2016}, the use of photoelastic methods on two-dimensional systems currently provide the best access to follow the dynamics of all particles and their vector contact forces \cite{daniels_photoelastic_2017, abed_zadeh_enlightening_2019, ramesh_developments_2021,JOSEPHANTONY2024108512}.

\begin{figure}
\centering
\includegraphics[width=0.3\linewidth]{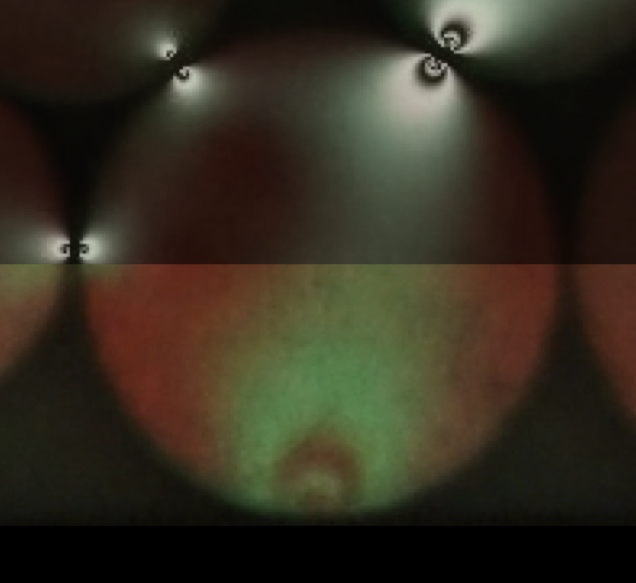}
\caption{A single grain imaged so that green light passes through a polariscope, revealing a photoelastic fringes due to forces at the perimeter of the grain. The top half of the image is overlaid with its pseudo-image (see section~\ref{dsolve}).}
\label{fig:grain}
\end{figure}

\begin{figure*}
    \centering
    \includegraphics[width=0.75\linewidth]{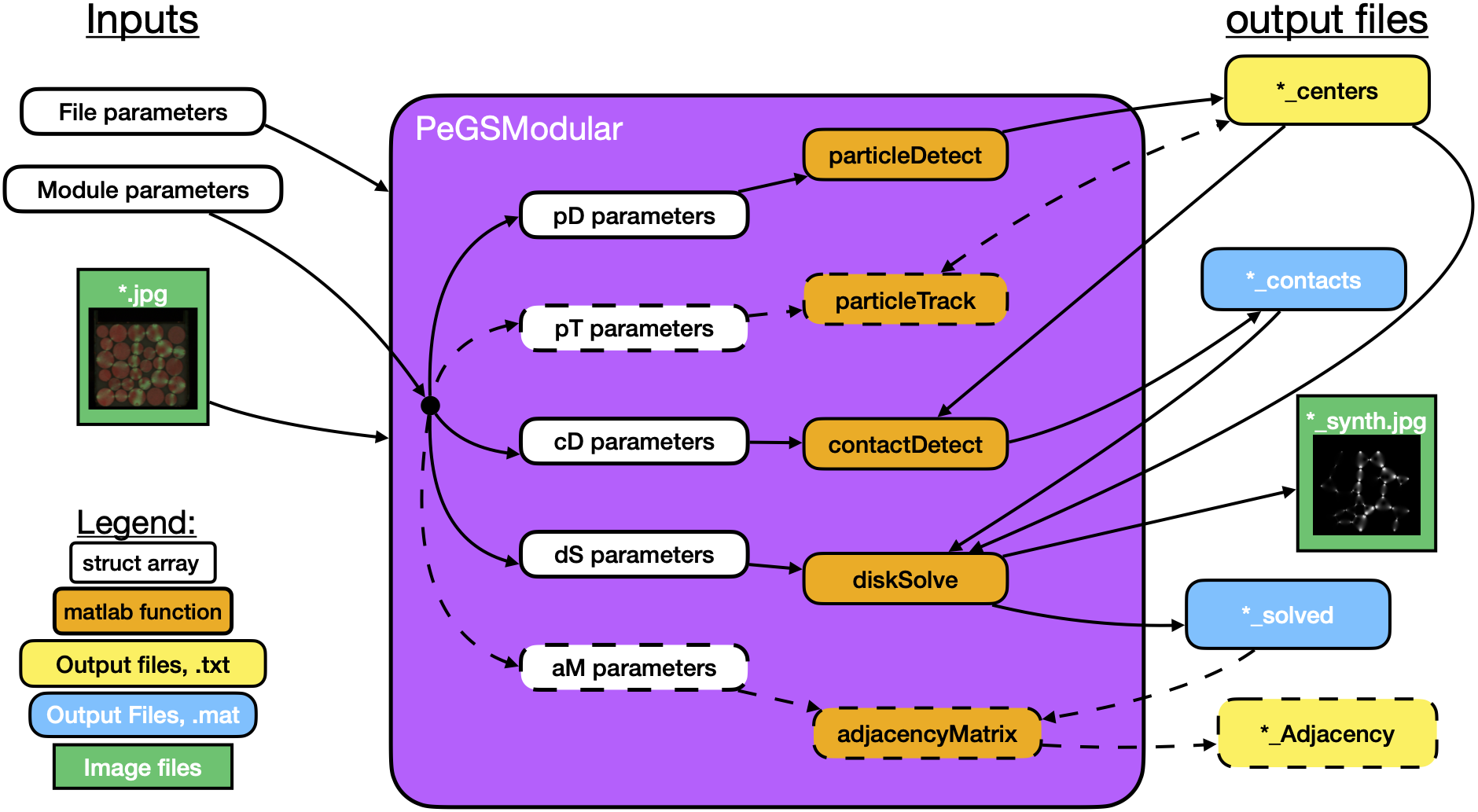}
    \caption{Schematic of the workflow and I/O of PeGS v.2.0.  Dashed lines indicate optional functions and their input/outputs.}
    \label{fig:overview}
\end{figure*}

Photoelastic materials are birefringent, and the strength of the birefringence is a function of the local stresses within the particle material. The difference between the refractive indices is linearly proportional to the local stress anisotropy.
As a result, a phase difference is introduced between the components of transmitted polarized light parallel to the principal stress axes. When viewed through a circular polariscope, the intensity of transmitted light is attenuated due to this phase difference is
\begin{equation}
    I=I_0\sin^2\left(\frac{\pi d \Delta n}{\lambda}\right)=I_0\sin^2\left(\frac{\pi d C\left(\sigma_1-\sigma_2\right)}{\lambda}\right)
    \label{stress-optic-law}
\end{equation}
where $I_0$ is the intensity of incident light, $d$ is the material thickness, $\lambda$ is the wavelength of the incident light$\sigma_1$ and $\sigma_2$ are eigenvalues of the local stress tensor, and $C$ is a wavelength-dependent material parameter known as the stress-optic coefficient. An example image of a birefringent particle is shown in Fig.~\ref{fig:grain}, with the polariscope image in the green channel. A useful set of resources on photoelasticity, curated by the community, is available at \url{https://photoelasticity.net} \cite{photoelasticity}. 

Since the pioneering work of Majmudar and Behringer~\cite{majmudar2005contact}, photoelastic images have been analyzed using an inverse method to determine the vector contact forces which gave rise to a particular fringe pattern.  The established procedure is to (1) identify particles in a polarigraph of the system; (2) identify possible contacts between particles; (3) estimate the magnitude of those contact forces; (4) generate the expected intensity across the particles, which we dub a {\it pseudo-image} and compare the pseudo-image to the polarigraph; and (5) iteratively refine the estimates of the contact forces to optimize agreement between the pseudo-image and the polarigraph, thus producing a measurement of those contact forces. A typical image and pseudo-image of a single grain are shown in Fig.~\ref{fig:grain}. This general approach has the advantage of providing a measurement of both the magnitude and direction of the contact forces (in contrast with methods that only measure the force magnitude), but requires high-resolution images with uniform lighting. 

Daniels et al.~\cite{daniels_photoelastic_2017} introduced an open-source software package, the Photoelastic Grain Solver, or PeGS (v.1.0), to carry out this inverse method analysis.
In November of 2023, a group of scholars who use photoelasticity in research of granular materials gathered for a workshop in Alexandria, VA, and identified improvements to PeGS as a primary goal of the community. Over the next year, PeGS was restructured to be more modular and to incorporate improvements to efficiency and precision. A subtle but important bug was identified and eliminated, and an organizational structure was developed to facilitate ongoing software development and support. In late 2024, we released PeGS v.2.0~\cite{lee_2024_15547001}, and report the updates and related resources here, including links to user-developed resources that form an important part of the development ecosystem.
As an example, the PeGS algorithm has been adapted to study cohesive granular materials by solving for tensile contact forces, details of which are reported in another article in this Proceedings 
\cite{naseer_png_preprint,naseer_cohesion}. 
PeGS v.2.0 makes it easier to disseminate and adapt software like this by establishing standardized functionality, input and output formatting, and data structures.

\section{PeGS v.2.0 structure and performance\label{sec:pegs2}}

PeGS v.2.0 is a series of Matlab functions using a shared data structure to apply the inverse method analysis~\cite{lee_2024_15547001}. The workflow mirrors the software structure, depicted as a flowchart in Fig.~\ref{fig:overview}. These functions can be easily invoked via the provided wrapper script ({\tt PeGSModular.m}) or a similar user-created wrapper program. An example application of this code may be found in another article in this Proceedings~\cite{clee_png_preprint}, and major improvements are enumerated in Table~\ref{tab:v2imp}. For a working example of how build the input arrays and call PeGSv2 via {\tt PeGSModular.m}, readers are encouraged to examine {\tt RunscriptSample.m} in the PeGSv2 github repository~\cite{lee_2024_15547001}.


\begin{table}[]
\begin{tabular}{|p{0.3\linewidth} | p{0.6\linewidth}|}
\hline\hline &Major improvements in PeGSv2\\ \hline
Structure&Functionalized rather than scripted\\
User I/O & Standardized struct arrays\\
Particle detection & User-defined input parameters and detection algorithm\\
Particle tracking & New feature to study dynamics \\
Regression & User-selectable algorithm \\
Customizeability & Straightforward to replace or insert functions \\
Development and support & Team of active developers with open team structure; code base remains open source, easily forked. \\
\hline
\end{tabular}
\caption{A table highlighting major improvements in PeGSv2.}
\label{tab:v2imp}
\end{table}

\subsection{Data structures \label{sec:datastruct}}

To preserve the modularity of the Matlab functions, we have standardized the input, output and data structures. If new modules are developed for community use, new developers must use the same formatting. Each module has three inputs, a structure containing file parameters, a structure containing module parameters, and a verbose flag. The file parameters structure includes information used to find input files and save output files, represented by green, yellow, and blue boxes in Fig.~\ref{fig:overview}. Each module has its own module parameter structure containing user set parameters needed to run the module. Verbose is a Boolean field that shows the user figures mid-use for quality control. The output of each function is a \textsc{true/false} flag to indicate successful completion of the current module along with module specific outputs saved in the designated directory.

\begin{figure}
\centering
\includegraphics[width=\linewidth]{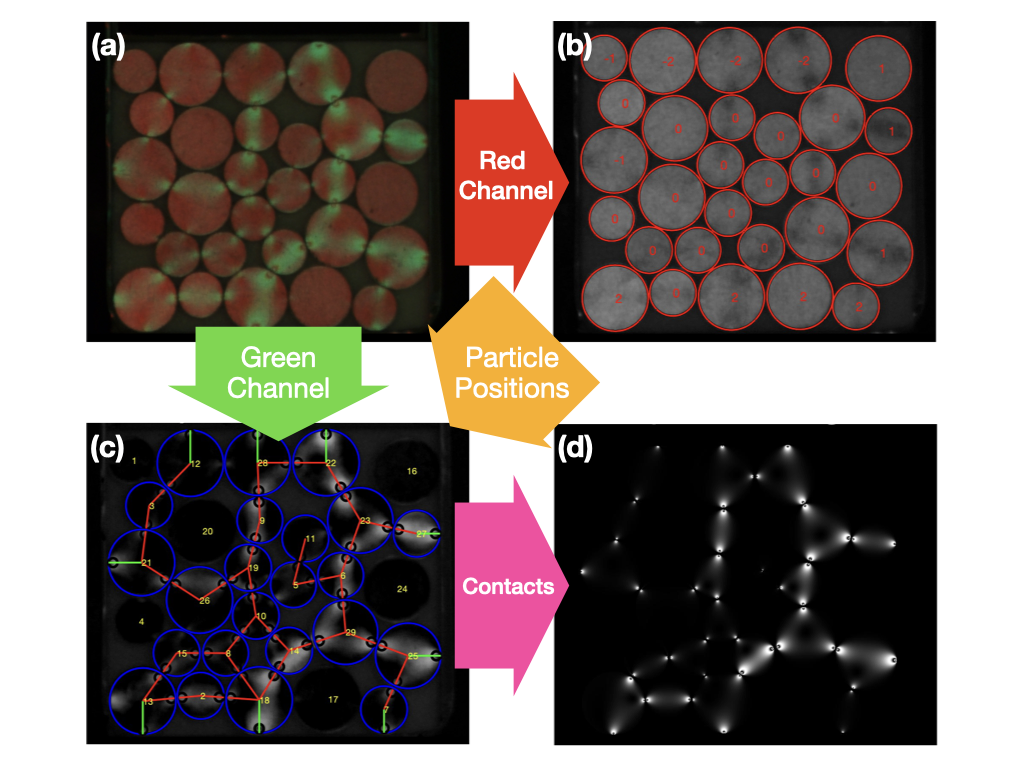}
    \caption{Images produced from the GitHub test dataset. From the red channel of \emph{(a)} the raw image, {\tt particleDetect.m} identifies \emph{(b)} all circular particles (the number label indicates bulk (0) or edge (non-zero) grains). The {\tt contactDetect.m} module identifies \emph{(c)} the contact network (red and green lines) using the particle locations and the green channel of the original image. The contact information is used to generate a \emph{(d)} Pseudo image with the {\tt diskSolve.m} module.}
    \label{fig:imagepipeline}
\end{figure}

\subsection{Image requirements \& pre-processing \label{sec:images}}

As a rule of thumb, we have found that a camera resolution that provides at least 50 pixels/particle is required for reliable fringe-fitting, and that uniform image contrast is quite important, particularly to precisely analyze the birefringence from polariscope images.
Uniform contrast requires uniform lighting of the experiment, but may also require image corrections due to optical artifacts, as implemented in Ref.~\cite{DongGithub}.
Polariscope images should use a monochromatic light source, as the optical response of the particles and optical elements of a polariscope are typically wavelength dependent. 
By default, PeGS expects input images to be 3-channel (RGB) color images which include a bright- or dark-field image of the particles (for particle detection) in the red channel, and a polariscope image in the green channel.
Currently, the blue channel is not used in default the PeGS analysis pipeline, but it may be incorporated in future updates, and can be used to capture additional information (e.g., particle orientation or fluorescent markers to facilitate more sophisticated tracking)  for use in user-created software.
 

\subsection{Particle Detection \label{sec:detect}}

Experimental images are loaded and processed individually using the function {\tt particleDetect.m}.
The particles are found using the MATLAB {\it imfindcircle} function, which uses a circle-finding algorithm based on the Hough transform. Users can set particular parameter values (e.g. sensitivity) within the function in order to optimize particle-detection. For each image, the output is a text file of particle positions, particle radii, and edge classification (i.e., whether a particle in the bulk or on a boundary). The update to this module streamlined the v.1.0 code, made setting parameters more apparent to end users, and updated the inputs and outputs to comply with the new standards.

\subsection{Contact Detection}
\label{cdetect}

The {\tt contactDetect.m} module finds contacts between particles in the system by taking each particle's image and calculating the local intensity gradient squared, or $G^2$, on the face of each particle. $G^2$ provides a semi-quantitative measure of the amount of stress on each particle~\cite{zhao_particle_2019}.
The module then takes the position data from particle detection to determine potentially neighboring pairs of particles. 
When combining the particles' positions and respective $G^2$ values against predefined user thresholds, the module determines the likely presence of contacts between particles in the system.
There are also special cases included to flag particle-wall and particle-free edge interactions.
Smaller forces require a more intricate algorithm than the $G^2$ values, for instance as has been used by \cite{lee_loading-dependent_2024} for more robust contact-detection.

\subsection{Disk Solve}
\label{dsolve}

The function {\tt diskSolve.m} solves for the vector of the interparticle forces acting at each contact. To do so, it calls the {\tt stress\_engine\_original.m} function, which takes a set of contact locations and forces as inputs, and produces the intensity field described in Eq.~(\ref{stress-optic-law}) by calculating the principal stress difference $(\sigma_1 - \sigma_2)$ at each point in the particle. 
The stress field is calculated assuming the particles are linearly elastic, flat, circular plates, as described in cite~\cite{daniels_photoelastic_2017}.
{\tt diskSolve.m} compares this "pseudo-image" to the experimental image of the photoelastic fringe pattern, and iteratively refines the estimated values for each contact force in order to optimize the agreement between the experimental photo and the pseudo-image.
In PeGS v.2.0, the user can set both the algorithm and tolerances used during the non-linear regression, and choose whether to impose force balance on each particle.

During inspection of the original PeGS code, we noticed \cite{mcmillan} that the stress-free boundary term arising from the transformation to a circular geometry was incorrect, producing nonphysical fringe patterns and potentially leading to erroneous values for the solved forces. In PeGS v.2.0, we provide an updated  formulation for the pseudo-image, accurately resolving the boundary terms. This correction has also been pushed in the most recent versions of the original PeGS v.1.0 software.


\subsection{Optional post-processing tools
\label{sec:options} }

The module {\tt particleTrack.m} performs  particle tracking by assigning a  unique ID to each particle in the ensemble, valid  across a series of images. We start by constructing a polygon around each particle center in the initial configuration. In each subsequent image, we detect the particle centers within the space demarcated by the polygons. The detected particle center is then assigned the same ID as the particle around which the polygon was constructed. This procedure is repeated for each strain step, where the polygons are constructed using particle center information from the preceding strain step. This approach provides a unique identity to each particle, which would otherwise randomly assigned by the \textit{imfindcircles} function in Matlab. 
This tracking functionality is new to PeGS2, and is one of the features that has made PeGS2 more useful for studying granular systems in flowing states.

There is an option available to provide a network visualization of the output. In the visualization loop, each particle's contacts are iterated over, and for each contact, a line is drawn from the particle to its contact point. The line's width and color are determined based on the normalized force exerted at that contact, where a wider and redder line indicates a higher force. 
For further applications requiring an adjacency matrix as an input (e.g. graph theory, topological data analysis), this is available as an optional output via the function {\tt adjacencyMatrix.m}.

\section{Notes from the development team \label{sec:devs}}

This project has benefited from community contributions, and we encourage those who are interested in becoming involved to get in touch with our team. Information on how to contact us, and to join an announcements mailing list, is available at \url{https://github.com/photoelasticity}.

\section{Acknowledgments}

This work was supported in part by National Science Foundation grants DMR-2046551, DMR-1846991, and DMR-2104986, SERB grant CRG/2022/003750, a Royal Society RFERE210264 Enhancement Grant,  a Moore Foundation Experimental Physics Investigator Initiative GBMF12236, and the Fulbright-Nehru fellowship program. We thank Eduardo Castellanos, Vir Goyal, and Jing Wang for feedback on pre-release versions of PeGS v.2.0.

~

\bibliography{PeGS2} 

\begin{thebibliography}{17}

\bibitem{dijksman_refractive_2017}
J.A. Dijksman, N.~Brodu, R.P. Behringer, Refractive index matched scanning and
  detection of soft particles, \textbf{88}, 051807 (2017).
  \doiwoc{10.1063/1.4983047}

\bibitem{hall_discrete_2010}
S.A. Hall, M.~Bornert, J.~Desrues, Y.~Pannier, N.~Lenoir, G.~Viggiani,
  P.~Bésuelle, Discrete and continuum analysis of localised deformation in
  sand using {X}-ray $\mu${CT} and volumetric digital image correlation,
  Géotechnique \textbf{60}, 315 (2010). \doiwoc{10.1680/geot.2010.60.5.315}

\bibitem{hurley_quantifying_2016}
R.C. Hurley, S.A. Hall, J.E. Andrade, J.~Wright, Quantifying {Interparticle}
  {Forces} and {Heterogeneity} in {3D} {Granular} {Materials}, Physical Review
  Letters \textbf{117}, 098005 (2016). \doiwoc{10.1103/PhysRevLett.117.098005}

\bibitem{daniels_photoelastic_2017}
K.E. Daniels, J.E. Kollmer, J.G. Puckett, Photoelastic force measurements in
  granular materials, Review of Scientific Instruments \textbf{88}, 051808
  (2017). \doiwoc{10.1063/1.4983049}

\bibitem{abed_zadeh_enlightening_2019}
A.~Abed~Zadeh, J.~Bares, T.A. Brzinski, K.E. Daniels, J.~Dijksman, N.~Docquier,
  H.O. Everitt, J.E. Kollmer, O.~Lantsoght, D.~Wang et~al., Enlightening force
  chains: a review of photoelasticimetry in granular matter, Granular Matter
  \textbf{21}, 83 (2019). \doiwoc{10.1007/s10035-019-0942-2}

\bibitem{ramesh_developments_2021}
K.~Ramesh, Developments in {Photoelasticity}: {A} renaissance (IOP Publishing,
  2021)

\bibitem{JOSEPHANTONY2024108512}
S.~{Joseph Antony}, Power of photo-stress analysis in unravelling the mechanics
  of granular materials and its applications in interdisciplinary research,
  Optics and Lasers in Engineering \textbf{183}, 108512 (2024).
  \doiwoc{10.1016/j.optlaseng.2024.108512}

\bibitem{photoelasticity}
Photoelasticity.net, \urlstyle{tt}\url{https://photoelasticity.net/}

\bibitem{majmudar2005contact}
T.S. Majmudar, R.P. Behringer, Contact force measurements and stress-induced
  anisotropy in granular materials, Nature \textbf{435}, 1079 (2005).
  \doiwoc{10.1038/nature03805}

\bibitem{lee_2024_15547001}
C.~Lee, L.~McCabe, B.~McMillan, A.~Naseer, D.~Xie, K.~Daniels, T.G. Murthy,
  K.~Nordstrom, T.~Brzinski, photoelasticity/pegs2: Initial release of pegsv2
  (2024), \urlstyle{tt}\url{https://doi.org/10.5281/zenodo.15547001}

\bibitem{naseer_png_preprint}
A.~Naseer, K.E. Daniels, T.G. Murthy, Extracting contact forces in cohesive
  granular ensembles, (Powders and Grains, accepted)  (2025).

\bibitem{naseer_cohesion}
A.~Naseer, {PeGS}\_for\_cohesion (2025),
  \urlstyle{tt}\url{https://github.com/nsrabrar/PeGS_for_Cohesion}

\bibitem{clee_png_preprint}
C.L. Lee, E.~Az\'ema, K.E. Daniels, The stress-force-fabric relation across
  shear bands, (Powders and Grains, accepted)  (2025).

\bibitem{DongGithub}
X.~Dong, {PeGS} pre- and post-process, \url{https://github.com/XD1729/PeGS2}
  (2025)

\bibitem{zhao_particle_2019}
Y.~Zhao, H.~Zheng, D.~Wang, M.~Wang, R.P. Behringer, Particle scale force
  sensor based on intensity gradient method in granular photoelastic
  experiments, New Journal of Physics \textbf{21}, 023009 (2019).
  \doiwoc{10.1088/1367-2630/ab05e7}

\bibitem{lee_loading-dependent_2024}
C.L. Lee, E.~Bililign, E.~Azéma, K.E. Daniels, Loading-dependent microscale
  measures control bulk properties in granular material: an experimental test
  of the {Stress}-{Force}-{Fabric} relation (2024).
  \doiwoc{10.48550/arXiv.2409.08140}

\bibitem{mcmillan}
B.~McMillan, S.~Dalziel, N.~Vriend, Validation and correction of photoelastic
  techniques for frictional granular systems, Measurement Science and
  Technology \textbf{36}, 055212 (2025). \doiwoc{10.1088/1361-6501/add48b}

\end{thebibliography}
\end{document}